# Deep learning for detecting pulmonary tuberculosis via chest radiography: an international study across 10 countries

**Authors**
Sahar Kazemzadeh[1†], Jin Yu[1†], Shahar Jamshy[1†], Rory Pilgrim[1], Zaid Nabulsi[1], Christina Chen[1], Neeral Beladia[1], Charles Lau[2], Scott Mayer McKinney[1], Thad Hughes[1], Atilla Kiraly[1], Sreenivasa Raju Kalidindi[3], Monde Muyoyeta[4], Jameson Malemela[5], Ting Shih[6], Greg S. Corrado[1], Lily Peng[1], Katherine Chou[1], Po-Hsuan Cameron Chen[1], Yun Liu[1*], Krish Eswaran[1], Daniel Tse[1*], Shravya Shetty[1‡], Shruthi Prabhakara[1‡]

**Affiliations**
[1] Google Health, Palo Alto, CA, USA
[2] Work done at Google via Advanced Clinical, Deerfield, IL, USA
[3] Apollo Radiology International, Hyderabad, India
[4] TB department, Center of Infectious Disease Research in Zambia, Lusaka, Zambia
[5] Sibanye Stillwater, Weltevreden Park, Roodepoort, South Africa
[6] Clickmedix, Gaithersburg, MD, USA
[†] equal contributions
[‡] equal contributions

*Address correspondence to: liuyun@google.com, tsed@google.com

# Abstract

**Purpose:** Tuberculosis (TB) is one of the top 10 causes of death worldwide and disproportionately affects low-to-middle-income-countries. Though the WHO recommends chest radiographs (CXRs) to facilitate TB screening efforts and the means of acquiring CXRs are generally accessible, expertise in CXR interpretation poses a challenge to broad implementation of TB screening efforts in many parts of the world. To help mitigate this challenge, we developed a generalizable deep learning system (DLS) to help detect active TB and compared its performance to 14 radiologists, from both endemic (India) and non-endemic (US) practice settings.

**Materials and Methods:** We trained a DLS using CXRs from 9 countries spanning Africa, Asia, and Europe. To improve generalization, we incorporated large-scale CXR pretraining, attention pooling, and semi-supervised learning via "noisy student". The DLS was evaluated on a combined test set spanning sites in China, India, US, and Zambia, with all positives confirmed via microbiology or nucleic acid amplification testing (NAAT). The India test set was independent of those used in training. Another independent test set from a mining population in South Africa was also used to further evaluate the model. Given WHO targets of 90% sensitivity and 70% specificity, the DLS's operating point was prespecified to favor sensitivity over specificity.

**Results:** Across the combined test set spanning 4 countries, the DLS's receiver operating characteristic (ROC) curve was above all 9 India-based radiologists (where TB is endemic), with an area under the curve (AUC) of 0.90 (95%CI 0.87-0.92). At the prespecified operating point, the DLS's sensitivity (88%) was higher than the India-based radiologists (mean sensitivity: 75%, range 69-87%, p<0.001 for superiority), and the DLS's specificity (79%) was non-inferior to these radiologists (mean specificity: 84%, range 78-88%, p=0.004). Similar trends were observed within HIV positive and sputum smear positive sub-groups, and in the additional South Africa test set. We additionally found that 5 US-based radiologists (where TB is not endemic) who also reviewed the cases were more sensitive but less specific than the India-based radiologists. The DLS was similarly non-inferior to this second cohort of radiologists at the same prespecified operating point. Depending on the simulated setting and prevalence, use of the DLS as a prioritization tool for NAAT could reduce the cost per positive TB case detected by 40-80% compared to the use of NAAT alone.

**Conclusion:** We developed a DLS to detect active pulmonary TB on CXRs, which generalized to patient populations from 5 different regions of the world, and merits prospective evaluation to assist cost-effective screening efforts in settings with scarce access to radiologists. Operating point flexibility may permit customization of the DLS to account for site-specific factors such as TB prevalence, demographics, clinical resources, and customary practice patterns.

# Introduction

Globally, 1 in 4 people are infected with *Mycobacterium tuberculosis*, and 5-10% of these individuals will develop active tuberculosis (TB) disease in their lifetime[1,2]. In 2019, the estimated TB mortality was 1.4 million, including 200,000 people who were human immunodeficiency virus (HIV) positive, and an estimated 2.9 million people who contracted TB were not formally reported due to a combination of underreporting, underdiagnosis, and pretreatment loss to follow up. Almost 90% of the active TB cases occur in a few dozen "high-burden" countries, many with scarce resources needed to tackle this public health problem.[3] The anticipated rising burden of drug resistant TB poses an increased threat to both endemic and non-endemic parts of the world.[4] Lastly, the COVID-19 pandemic that has caused devastation around the world has also disrupted efforts to combat TB: globally, 21% fewer (1.4 million) people received care for TB in 2020 than in 2019.[5]

In the past decade, there has been steady global support to combat this health crisis through the World Health Organization (WHO)'s End TB Strategy, the United Nations (UN)'s Sustainable Development Goals, and the Global Fund to fight AIDS, TB and malaria.[6] Cost effective pulmonary TB screening using CXR has the potential to increase equity in access to healthcare, particularly in difficult-to-reach populations.[7] In light of high patient volumes and limited access to timely expert interpretation of CXRs in many regions, there has been active research into using artificial intelligence to screen with a CXR followed by a corroborating diagnostic test;[8–20,21] Such artificial intelligence-based triaging followed by GeneXpert testing for a confirmatory diagnosis was shown to be cost-effective compared to GeneXpert alone, and also substantially increased patient throughput.[13] As part of their recently-published 2021 guidance, the WHO evaluated three independent computer-aided detection (CAD) software systems, and determined that the diagnostic accuracy and performance of CAD software was similar to human readers.[7,9,13,17] Given the scarcity of experienced readers, as an alternative to human interpretation of CXR, the WHO now recommends CAD for both screening and triage in individuals 15 years or older.[7] However, the WHO emphasized the importance of using a performant CAD system that has been tested on a population that is representative of the target population.

In this study, we developed a deep learning system (DLS) to interpret CXRs for imaging features of active TB. Developing a universal TB classifier can be challenging - not only due to the array of potential imaging features - but because prevailing imaging features, severity of disease at presentation, and prevalence of TB and HIV can differ broadly on locale. Therefore, we validated our DLS using an aggregate of datasets from China, India, US, and Zambia that together reflect different regions, race/ethnicities, and local disease prevalence. We evaluated the DLS under two conditions: having a single prespecified operating point across all datasets, and when customized to radiologists' performance in each locale. As diagnostic performance may be influenced by disease prevalence, we compared the DLS with two different cohorts of radiologists: one based in a TB-endemic region (India) and one based in a TB non-endemic region (United States). An analysis of HIV positive and sputum smear positive subgroups was also performed. Finally, we estimate cost savings for using this DLS as a triaging solution for nucleic acid amplification testing (NAAT) in screening settings.

## Methods

### Datasets

For this work, we leveraged de-identified CXR images from multiple datasets spanning 9 countries for training and 4 countries for validating the DLS, for a total of 10 countries (Table 1). Our DLS was trained using 160,187 images from Europe[22] (Azerbaijan, Belarus, Georgia, Moldova, Romania), India, and South Africa, and tuned using 3,258 images from China[16,23,24], India, and Zambia[25]. Additionally, we used 550,297 images[26,27] for pretraining purposes (10,310 of which overlapped with the train sets and none of which overlapped with the tune sets), and 138 images with labeled lung segmentation masks from the US dataset for training and tuning of the lung cropping model. We then validated the DLS using 1,262 images from China[16,23], India, US[16,23], and Zambia, and 1040 images from South Africa, using 1 image per patient in both cases. Additional details including inclusion/exclusion criteria, enrichment, and reference standard are presented in Supplementary Figure S1 and Supplementary Table S1. This retrospective study was approved by the respective Ethics Committee or Institutional Review Board at each participating institution and all data were de-identified prior to transfer.

### Reference standard for TB status

For all test and tune datasets, the positive TB status were confirmed via microbiology (sputum culture or sputum smear) or NAAT (GeneXpert MTB/RIF, Cepheid, Sunnyvale, CA); see Table 1 and Supplementary Table S1. On the train datasets, the reference standard varied due to site-specific practice differences and data availability, including microbiology, radiologist interpretation, clinical diagnoses (based on medical history and imaging), and NAAT.

## Deep Learning System

### Deep Learning System Architecture

We developed a DLS to detect evidence of active pulmonary TB on CXRs. The system consists of three modules: a lung cropping model for identifying a bounding box spanning the lungs, a detection model for identifying regions containing possible imaging features of active tuberculosis (nodules, airspace opacities with cavitation, airspace opacities without cavitation, pleural effusion, granulomas, and fibroproductive lung opacities), and a classification model that takes the output from both the segmentation model and the detection model to predict the likelihood of the CXR being TB positive (Figure 1).

For the lung cropping model, we used Mask RCNN[28] with a ResNet-101-FPN[29] feature extractor to train for both pixel-level segmentation and bounding boxes as outputs. We then cropped each CXR using the model's output bounding box enclosing the lungs as the input for the classification model. For the detection model, we used a Single Shot MultiBox Detector (SSD)[30] to create bounding boxes around potential TB-relevant imaging features. Based on the predicted bounding boxes, a probabilistic attention mask was calculated as the final pooling layer in the classification model. For the classification model, we combined an EfficientNet-B7[31] pre-trained on classifying CXRs as normal or abnormal[27] with an attention pooling layer and a fully-connected layer. The attention pooling layer utilizes the probabilistic attention mask generated from the detection model to perform a weighted average of the feature maps before feeding to the final fully connected layer. The classification model classifies CXRs into 1 of 3 classes: TB-positive, TB-negative but abnormal, and normal. We took the prediction score for TB-positive class as the output prediction for all TB-related analysis.

## Deep Learning System Training

Training the individual components of the DLS described above is a multi-step process (Figure 1). First, we trained the lung cropping model using lung segmentation masks from the US dataset of 138 images with 80% used for training and 20% used for tuning. To train the detection model, radiologists annotated 9,871 bounding boxes around TB-indicative abnormalities (nodules, airspace opacities with cavitation, airspace opacities without cavitation, pleural effusion, granulomas, lymphadenopathy, and fibroproductive lung opacities). Both the detection and classification models were trained using the Europe dataset and the two India train datasets. Due to the limited amount of labeled data, we used the noisy-student[32] semi-supervised learning approach to leverage a much larger set of unlabeled data. Specifically, we obtained "noisy" TB labels by running inference using the initial version of the DLS on the South Africa train dataset with more than 150,000 unlabeled CXRs. These data with generated labels were combined with the original dataset to train 6 classification models, which were then ensembled by taking the mean of the scores.

For the detection model, we used a dropout keep probability of 0.99, and augmentation included random cropping, rotation, flipping, jitter on the bounding boxes, multi-scale anchors, and a box matcher with intersection-over-union. For the classification model, we applied dropout, with a dropout keep probability of 0.5. Furthermore, we applied data augmentation such as horizontal flipping, random shears, and random deformations. All hyperparameters were selected based on empirical performance on the tune sets. Training was done using TensorFlow on third-generation tensor processing units with a 4x4 topology. All images were scaled to 1024 x 1024 pixels, and image pixel values were normalized on a per-image basis to be between 0 and 1.

## Model Selection

For model selection (checkpoint selection and other hyperparameter optimization), we selected models to maximize the area under the receiver operating characteristic curve (area under ROC curve, or AUC) corresponding to the range of radiologists' sensitivities in the tune sets. This approach was used to help explicitly select models that were performant across the range of radiologist sensitivities, instead of potentially optimizing for ranges that were beyond the scope of customary clinical practice.

## Comparator Radiologist Reviews

In order to gauge the performance of the DLS across datasets containing cases of different levels of difficulty, all test set cases were reviewed by a team of radiologists, whose performance not only served as a baseline for comparison, but also as an indirect indicator of difficulty level. As the performance characteristics of radiologists accustomed to practice in endemic vs. non-endemic settings can vary, this team consisted of two cohorts of radiologists (10 India-based consultant radiologists and 5 US-based board-certified radiologists). The India-based radiologists had an average of 6 years of experience (range 3-9), while the US-based radiologists had an average of 10.8 years of experience (range 3-22). These radiologists were provided with both the image and additional clinical information about the patient when available (age, sex, symptoms, and HIV status), whereas the DLS was blinded to these information. For each image, the radiologists labeled it for the presence/absence of TB, other pulmonary findings, and optionally whether there were any minor technical issues visible on the image. The tune sets were labeled similarly.

## Statistical Analysis

Our primary analyses compared the performance of the DLS with that of the India-based radiologists on the pooled combination of 4 test datasets. To support comparisons with the binary judgments of experts, we thresholded the DLS's continuous score using an operating point of 0.45, chosen based on an analysis of the tune datasets (conducted prior to evaluating the DLS on any of the test sets). We tested for noninferiority of sensitivity and specificity, both with a 10% absolute margin. To account for correlations within case and within radiologist, we used the Obuchowski-Rockette-Hillis procedure[33,34] configured for binary data[35] and adapted to compare readers with the standalone algorithm[36] in a noninferiority setting[37]. A p-value below 0.0125 was considered significant for the primary analyses (a conservative one-sided alpha of 0.025 was halved for a Bonferroni correction for 2 tests). Subsequent superiority testing was prespecified if non-inferiority was met, which does not require multiple testing correction.[38]

Prespecified secondary analyses included per-dataset subgroup analysis for ROC; sensitivity and specificity at the prespecified operating point; operating points corresponding to the WHO thresholds; matched sensitivity/specificity analysis on a per-dataset and per-radiologist level; comparisons of the India-based and US-based radiologists, and comparison of the DLS to the US-based radiologists. Additional secondary analyses were on subgroups based on HIV status, images flagged by the reviewing radiologists to have minor technical issues, demographic information, and symptoms (work in progress). Exploratory subgroup analysis based on sputum smear was also conducted. Unless otherwise specified, 95% confidence intervals (CIs) were calculated using the bootstrap method with 1000 samples.

## Labeling for non-TB Pulmonary Findings

To understand the performance of the "abnormality" detector in the DLS, we additionally labeled the India test dataset for any actionable abnormal CXR findings. Each case was reviewed by 3 US-based radiologists. Because follow up testing such as a repeat CXR or computed tomography were not available, "ground truth" was based on how many radiologists indicated the presence of an abnormal finding: at least 1 of 3, at least 2 of 3, and all 3 of 3.

## Cost Analysis

Finally, we simulated the potential cost savings of using our DLS as a TB screening intervention. Recent studies have estimated the overall cost for subsidized GeneXpert to be about US$13.06 per test, including equipment, resources, maintenance, and consumables.[13] The cost to acquire a single digital CXR was estimated to be US$1.49, including equipment and running costs, but not radiology interpretation.[13,39] In our simulation, the DLS is used for initial TB screening, and patients who meet the threshold (based on our prespecified operating point) proceed with GeneXpert testing. The expected total GeneXpert testing cost was computed using the prevalence, sensitivity, and specificity to get DLS-positive rates and multiplying by the cost of GeneXpert. The total expected cost included both this testing cost for DLS-positive patients and the cost of CXR screening for all patients. Finally, we divided the total cost by the number of true positive TB cases caught to derive the cost per positive TB case.

We then analyzed the effect of prevalence on the cost, which makes the simplifying assumption that there are no changes in case severity or other factors that may affect DLS performance.

## Results

DLS performance was first evaluated on a combined test dataset incorporating a diverse population representing multiple races and ethnicities, drawn from 4 countries: China, India, US, and Zambia (Table 1). Among a total of 1,262 images from 1,262 patients, there were 217 TB cases based on positive culture or GeneXpert. DLS development (training and tuning) and operating point selection was conducted on the tune datasets, independently of the test datasets. Patient sources for these 4 datasets included TB referral centers, outpatient clinics, and active case finding. The India test dataset was from a site independent of those used in development. An independent dataset from South Africa comprising a mining population served as an additional test set.

### DLS Performance

In our combined test dataset across 4 countries, the DLS achieved an AUC of 0.90 (Figure 2A, Table 2). To contextualize the model's performance and better understand the case spectrum, we obtained radiologist interpretations for the same cases from two cohorts of radiologists: radiologists based in India, a country where TB is endemic, and radiologists based in the US. One India-based radiologist was found to have a rate of flagging positives (and consequently sensitivity) substantially below the others (Supplementary Figure S2) and so was excluded from subsequent analyses to avoid under-representing radiologist performance. The DLS's ROC curve was above the performance points of all 9 remaining India-based radiologists (Figure 2A).

Our prespecified primary analyses involved comparisons of the DLS at a prespecified operating point (0.45) with India-based radiologists. The DLS's sensitivity (88%, 95% CI 83-94%) was higher than (superiority test was conducted if non-inferiority passed, see Methods) the India-based radiologists (median sensitivity: 74%; IQR: 72-76%), $p<0.001$. At the same operating point, the DLS's specificity (79%, 95%CI 75-82%) was similarly non-inferior to the India-based radiologists (median specificity: 86%; IQR: 81-87%), $p=0.003$.

### Comparison of India-based and US-based Radiologists

While both India-based and US-based radiologists had sensitivities and specificities that tracked closely and slightly below the ROC curve of our model, the conservativeness with which the two groups of radiologists called cases as positive for TB appeared to differ (Figure 2A-C, Supplementary Figure S2). India-based radiologists appeared to be more specific but less sensitive than US-based radiologists, who had a median sensitivity of 84% (IQR 76-86%) and a median specificity of 71% (IQR 67-81%). The DLS's sensitivity and specificity remained comparable to the US-based radiologists (p-value for non-inferiority: 0.022 for sensitivity; 0.018 for specificity).

### Per-dataset Analysis

Next, we conducted subgroup analysis on a per-dataset level (Table 2 and Figure 2B). The China and US datasets were similarly-constructed case-control datasets, with normal CXRs selected to match the TB positive CXRs. On these two datasets, while the India-based radiologists achieved high specificity (96-99%), their sensitivity was lower (53-65%) compared to in the combined dataset. At the prespecified operating point, both the DLS's sensitivity and specificity were non-inferior to the radiologists in both datasets (p<0.001 for all 4 comparisons). In the India dataset, which consisted of TB presumptive patients identified in a tertiary hospital, the DLS was similarly non-inferior in both sensitivity and specificity (p<0.001 for both). In the Zambia dataset, which was taken from a trial[25], NAAT was associated with cases where the CAD4TB system had flagged an abnormal CXR, resulting in substantial enrichment for CXR-abnormal TB-negatives. In this dataset, at the prespecified operating point, the DLS was non-inferior for sensitivity (p<0.001) but not for specificity (p=0.504), though 8 of 9 India-based radiologists were below the ROC curve.

In addition to the 4 datasets above, we evaluated the DLS on another independent dataset from a mining population in South Africa (Figure 2D and Supplementary Table S2). The ROC curve of the model was above all but 1 radiologist. At the same prespecified operating point as the other datasets, the DLS was non-inferior both in terms of sensitivity and specificity to both India-based and US-based radiologists (p<0.05 for all). At a higher (lower sensitivity) operating point selected based on the South Africa tune datasets, the DLS was again non-inferior in both sensitivity and specificity compared to the India-based radiologists, but had higher specificity (p=0.012) at the cost of not being non-inferior in sensitivity (p=0.571).

### Inter-dataset Comparisons

To better understand inter-dataset differences, histograms of DLS prediction scores were plotted separately for TB positive and TB negative cases for each dataset (Figure 3). The distribution of DLS scores for both TB positive and TB negative cases remained similar across the China, India, and US datasets (Supplementary Table S5). However, there was a higher proportion of TB-negative cases with high DLS scores in the Zambia dataset. This appears to have been a consequence of first-round CAD screening of the Zambia dataset which censored many normal-appearing CXRs, resulting in a more challenging dataset with a relative paucity of normal CXRs.

### Matched Performance to Radiologists

To facilitate comparisons despite the wide range in radiologists' sensitivities and specificities, both across datasets and readers, we next conducted a matched analysis by shifting the DLS's operating point on a per-dataset level to (1) compare sensitivities at mean radiologist specificity, and (2) compare specificities at mean radiologist sensitivity. These analyses were done separately for the India-based radiologists and US-based radiologists, for a total of 16 analyses (4 datasets * 2 comparator radiologist group * matching sensitivity/specificity) and presented in Table 3. The DLS had non-inferior performance in 15 out of these 16 analyses (p<0.05 for 15 and p=0.068 for the remaining).

Next, we adjusted the DLS's operating point to match each individual radiologist's sensitivity and specificity, focusing on the two larger datasets (India, Zambia) to improve statistical power. With 14 radiologists, 2 datasets, and matching sensitivity/specificity, this amounted to 56 analyses (Supplementary Table S3). In 50 of these analyses, the DLS was non-inferior (p<0.05), with 4 of the 6 non-passing tests in the enriched Zambia dataset and in comparison with US-based radiologists.

### WHO Target Sensitivity and Specificity

The WHO "target product profile" for a TB screening test recommends a sensitivity ≥90% and a specificity ≥70%. To further understand the performance of the DLS, we conducted matched performance analysis, similarly to the radiologist-matched analysis above. At 90% sensitivity, the DLS had a specificity of 77% on the combined dataset; and at 70% specificity the DLS had a sensitivity of 93%, both of which met the recommendations. This remained true in the China, India, and US datasets, but not in the enriched Zambia dataset (Table 4).

### Subgroups by HIV Status

We next considered subgroups based on HIV status where available (this included most patients in the Zambia dataset). The DLS found the HIV-positive subgroup more challenging than the HIV-negative subgroup (DLS AUC: 0.81 vs 0.92), and a similar lowering of sensitivity and specificity were observed for the radiologists (Figure 2C). However, the DLS remained comparable to the radiologists in both subgroups, notably despite the DLS not having access to the HIV status as the radiologists did.

### Subgroups by Sputum Smear

Sputum smear microscopy is fast, inexpensive, and specific for *Mycobacterium tuberculosis*. Despite the low sensitivity, it is still used for rapid diagnosis in resource limited settings.[40] We evaluated the performance of our model on this subset using our Zambia dataset, and evaluated the DLS's sensitivity for TB-positive cases with different sputum smear results. Although this subset was small with only 12 smear positive and 14 smear negative patients, at our prespecified operating point, our model had 100% sensitivity for smear-positive TB-positive patients and 71% sensitivity in smear-negative TB-positive patients.

### Subgroups by Demographic Information, TB History, and Symptoms

We also evaluated the DLS in subgroups based on age and sex (Supplementary Figures S3-4). The DLS's AUC varied between 0.86 to 0.97 within these subgroups, with similar trends of the ROC curve remaining higher than almost all of the radiologists.

### Sensitivity Analysis to Technical Issues

During case review, radiologists could indicate that images had technical issues that hindered confident interpretation. As the number of radiologists who indicated such issues grew from 0 to 1 to 2, the DLS AUC decreased, from 0.99 to 0.91 to 0.82 (Supplementary Figure S5). When grouping these images using a cumulative approach (i.e., "1 or more", "2 or more"), the trends were similar. However, of the 45 images where 3 or more radiologists indicated a technical issue, the AUC trend reversed to 0.90, though the confidence intervals grew. As may be expected, the radiologists' sensitivities and specificities moved in a similar manner for cases they had indicated issues with.

### Detecting non-TB Pulmonary Findings

We further evaluated the "abnormality" detector in the DLS on the India test set, using labels provided by 3 US-based radiologists as the "ground truth" (Methods). The DLS was then evaluated using this ground truth in 2 ways. First, we used the entire India test set and defined a positive case as either being TB-positive or having another abnormal CXR finding. Depending on how many radiologists indicated the presence of the abnormality, the DLS's AUC ranged from 0.80 to 0.96 (Supplementary Figure S6). Second, using only TB-negative cases, we defined a positive case as having any abnormal CXR finding, and plotted the ROC for the DLS's "non-TB abnormality" prediction alone. The AUC of the DLS ranged from 0.71 to 0.85, though with wider confidence intervals (Supplementary Figure S6).

### Cost Analysis

Finally, in our analysis of potential cost savings, we simulated a workflow where patients only proceed to GeneXpert testing if they are flagged as positive by the DLS. This workflow has a reduced overall sensitivity (though still exceeding the WHO target of 90%), but substantially reduces cost via a lower number of confirmatory tests being conducted and thus improves cost effectiveness as measured by cost per positive TB case detected. We then simulated the cost of using the DLS performance on the India dataset (94% sensitivity and 95% specificity), the WHO target performance (90% sensitivity and 70% specificity), a lower-specificity device (90% sensitivity and 65% specificity), and GeneXpert only (no CXR). Based on the performance on the India dataset, as prevalence decreases from 10% to 1%, the cost per positive TB case detected increased substantially, and the cost savings compared to using GeneXpert alone increased from 73% to 82% (Figure 5). The corresponding cost savings at low prevalence is not as profound when simulating the WHO target (47% to 53%) and a lower-specificity device (42 to 48%).

## Discussion

In order to achieve the long term public health vision of global elimination of TB, there is a pressing need to scale up identification and treatment in resource-constrained settings. The recently-released 2021 WHO consolidated guidelines stated that CAD technologies had the potential to "increase equity in the reach of TB screening interventions and in access to TB care." They also emphasized the importance of using a high-performing CAD that was tested on CXRs drawn from a representative population for the corresponding use case.[7,41] We have developed a DLS using data from 9 countries and validated the DLS in 5 countries, together covering many of the high-TB-burden countries and a wide range of race/ethnicities and clinical settings. In this combined international test dataset, the DLS's pre-specified operating point demonstrated higher sensitivity and non-inferior specificity relative to a large cohort of India-based radiologists. The development of a DLS with robust performance across a broad spectrum of patient settings has the potential to equip public health organizations and healthcare providers with a powerful tool to reduce inequities in efforts to screen and triage TB throughout the world.

When considering each dataset individually, the DLS's performance was excellent in two commonly-used case-control datasets from China and US, and generalized well to an external validation set in India. Moreover, the DLS's performance was maintained in the enriched Zambia dataset, which was filtered by another CAD device. Since many images that were considered radiologically clear were excluded from this dataset, the difficulty of triaging the remaining cases was likely increased. The DLS also performed well when radiologists indicated minor technical issues with the image, indicating robustness to real-world issues.

In addition to performing well in different countries with a wide range of race/ethnicities, the model was also comparable to radiologists in important subgroups. First, HIV infection increases the risk of active

TB disease up to 40 fold compared to background rates.[1,42] Patients with HIV-associated pulmonary TB often have an atypical presentation on CXR, making them more difficult to screen.[43] Thus, the fact that the DLS's detection performance remained comparable with radiologists in HIV-positive patients is reassuring. Second, sputum smear has a fast turnaround and a low cost, leading to its importance in resource-limited settings despite having limited sensitivity. Though the subgroups were small, the DLS was able to identify all sputum-positive cases and remained accurate on sputum-negative cases. The fact that the use of the DLS should not lead to missing cases that would otherwise be detected by a relatively accessible procedure is comforting, but will need to be further validated in a larger population. We further verified that the DLS remained comparable to the radiologists in important populations based on demographic information, for patients without a prior history of TB, and subgroups based on symptoms including WHO's recommended 4-symptom screen: cough, weight loss, fever, or night sweats. Importantly, our test set comprising a gold mining population in South Africa is supportive evidence for the DLSs potential to help with this subgroup recommended by the WHO for systematic screening. Finally, the DLS was able to accurately detect other non-TB abnormalities that were identified by radiologists. Such a capability resolves one of the drawbacks that traditional CAD systems were noted to have by the WHO: that unlike human readers, the CAD systems could not simultaneously screen for pulmonary or thoracic conditions.

Although NAAT, such as GeneXpert have high positive predictive value, many populations are unable to derive the broadest possible benefit from such tests because of their higher relative per-unit cost. However, if coupled to an inexpensive but relatively sensitive first-line filter like CXR (i.e., only cases screening positive on CXR are tested using NAAT), the benefits of NAAT could effectively benefit a larger population due to more targeted use. Two-stage screening strategies of this type would traditionally be intractable in many locales because settings with constrained access to NAAT often also lack providers trained to reliably interpret CXRs for TB-related abnormalities. In these settings, in accordance with current WHO guidelines, a robust performing CAD can increase the viability of this strategy by serving as an effective alternative to human readers. Our cost analysis of this two-stage screening workflow using the DLS suggests that it has the potential to provide 40-80% cost savings at 1-10% prevalence. The cost savings increases further as prevalence falls, which is an important financial consideration in disease eradication.

Our comprehensive analysis with a large cohort of radiologists also revealed several important subtleties. First, irrespective of practice location, radiologists demonstrated a wide range of sensitivities and specificities. For example, even among our 9 India-based radiologists, sensitivities spanned a 18% range (69-87%). This variability is documented in the literature, with clinical experience being a potential contributing factor.[7,44–46] However, this means that direct, single-operating-point comparisons with any individual radiologist can be difficult to interpret without matching the DLS operating point to either the sensitivity or specificity of that reader. Second, radiologists practicing in India were generally more specific and less sensitive than those practicing in the US. This may partially be due to practice patterns: in India where TB is endemic, radiologists' calls need to be highly specific to avoid testing an overwhelming number of patients. By contrast, in the US where TB is relatively rare and the goal is to avoid outbreaks, radiologists are incentivized to make calls that are highly sensitive at the expense of specificity. The wide range in performance between individuals and across practice locations suggests that future CAD for TB studies will likely need to take into account the practice locations of the comparator radiologists, and ensure that a sufficient number of radiologists are recruited to represent the natural variability.

Remarkably, despite variability in individual radiologists' sensitivity and specificity, their performance tracked closely with the DLS's ROC curve, with the clearest evidence of this trend seen in the enriched Zambia dataset which exhibits a marked rightward shift toward lower specificity. This suggests that the inherent advantageous ability of the DLS to provide continuous "scores" as output for thresholding can likely help individual sites customize the triggering rate to their local practice patterns and resource needs, while trusting that the customized operating point has a similar effect to calibrating the "conservativeness" of a radiologist. As suggested by the WHO, this ability to calibrate the operating point may be critical over time even for the same population, as prevalence and disease severity changes over time. Statistical methods to tune operating points for each dataset and to detect when operating points should be updated over time may be useful in this regard and represent an important direction of future work for real-world use cases.

Our study has limitations. First, this study was retrospective and prospective validation will be needed to better understand challenges in integrating into real-world workflows and to rigorously determine the TB status via mycobacterial culture for all patients. Second, as highlighted by the WHO, TB screening often happens in a broader context of patient care; patients can present with symptoms suspicious of TB but have other pulmonary or thoracic conditions instead. The subgroup analyses presented here were also not fully comprehensive due to the lack of important variables (such as HIV status and symptoms) in several datasets. As our datasets all had relatively high prevalence, we will need to evaluate its performance in populations with lower prevalence. Finally, the cost analysis is a simulation that makes simplifying assumptions such as DLS performance being unaffected by prevalence changes. In practice, prevalence changes may be associated with case severity, which may affect DLS sensitivity or specificity.

## Conclusion

In this study, we developed a DLS and demonstrated its generalization via international test datasets spanning 5 countries encompassing a wide range of race/ethnicities: China, India, US, and Zambia. At a uniform, prespecified operating point, the DLS had significantly higher sensitivity while maintaining non-inferior specificity compared to 9 India-based radiologists. When compared with US-based radiologists who were more sensitive but less specific, the DLS was non-inferior in both sensitivity and specificity. The DLS further meets the WHO targets when matching to either 90% sensitivity or 70% specificity. The DLS may be able to facilitate TB screening in areas with scarce radiologist resources, and merits further prospective clinical validation.

## Acknowledgements

The authors thank the members of the Google Health Radiology and labeling software teams for software infrastructure support, logistical support, and assistance in data labeling. For tuberculosis data collection, thanks go to Sameer Antani, Stefan Jaeger, Sema Candemir, Zhiyun Xue, Alex Karargyris, George R. Thomas, Pu-Xuan Lu, Yi-Xiang Wang, Michael Bonifant, Ellan Kim, Sonia Qasba, and Jonathan Musco. The train dataset from Europe/India was obtained from the TB Portals (https://tbportals.niaid.nih.gov), which is an open-access TB data resource supported by the National Institute of Allergy and Infectious Diseases (NIAID) Office of Cyber Infrastructure and Computational Biology (OCICB) in Bethesda, MD. These data were collected and submitted by members of the TB Portals Consortium (https://tbportals.niaid.nih.gov/Partners). Investigators and other data contributors that originally submitted the data to the TB Portals did not participate

# Figures

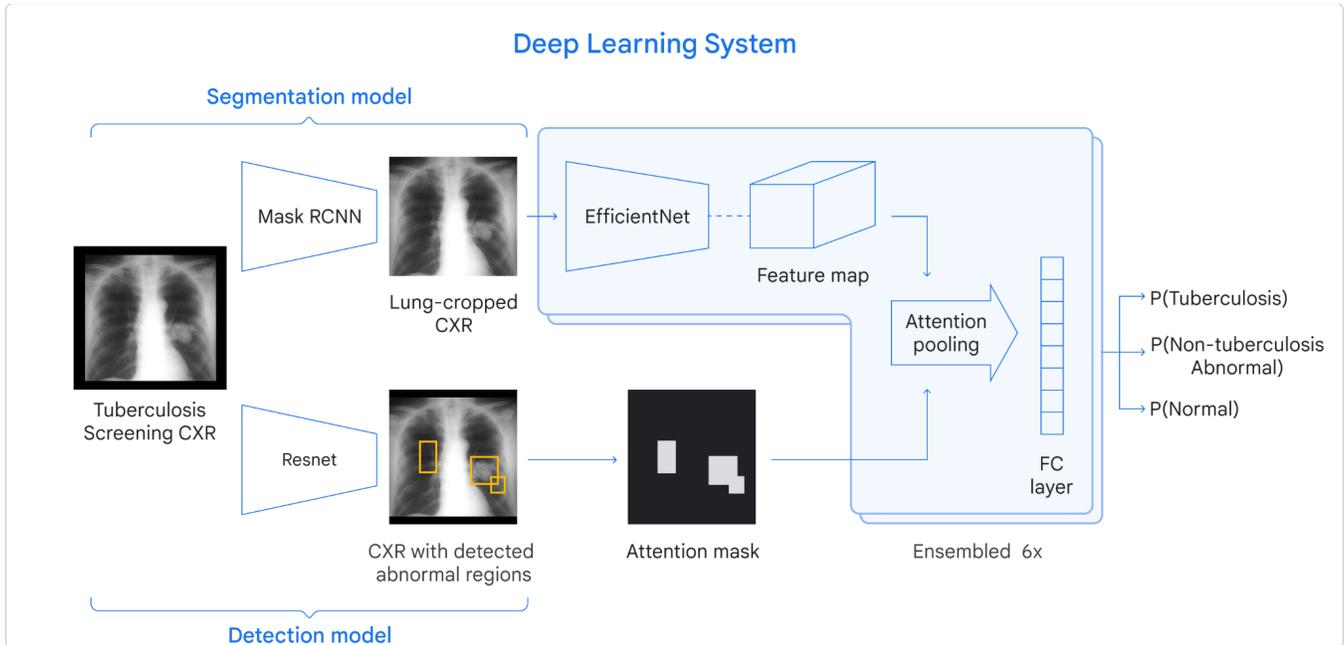

**Figure 1. Overview of our deep learning system (DLS).** The system consists of three modules: a lung segmentation model to specifically crop the lungs, a detection model to identify regions of interest, and a classification model that takes the output from the other two models to predict the likelihood of the CXR being TB positive. The large-scale abnormality pretraining and noisy student semi-supervised learning used to train these modules are not visualized here.

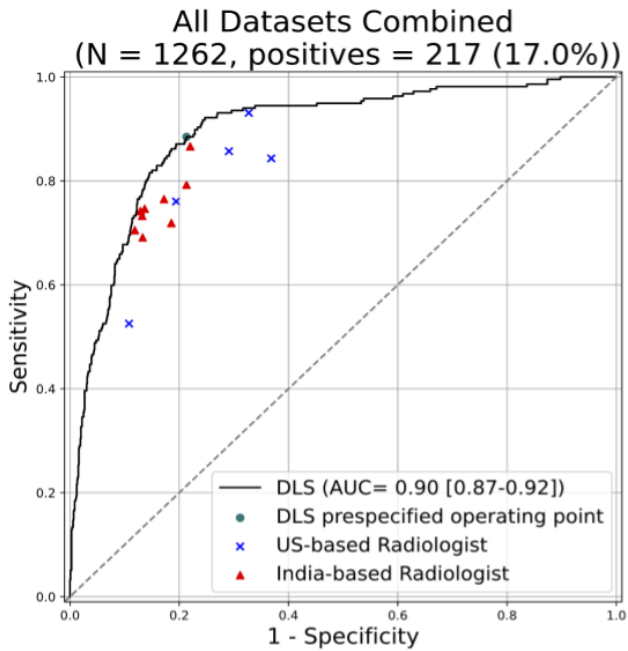
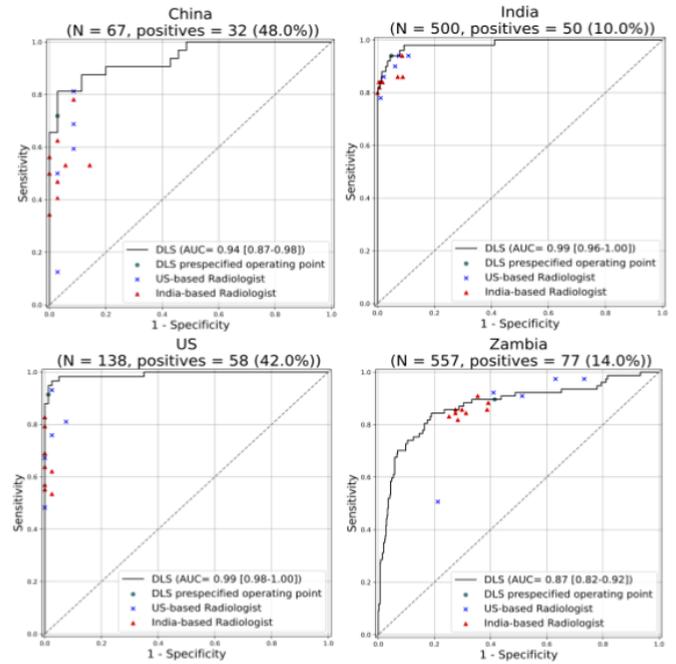
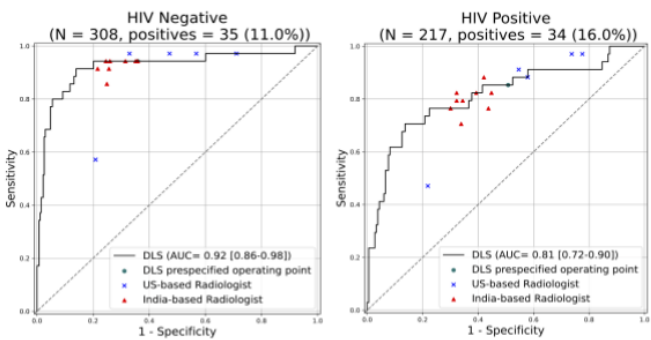
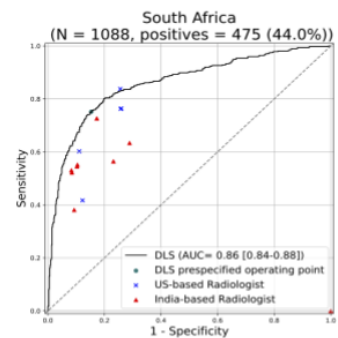

**Figure 2.** Receiver operating characteristic (ROC) curves for the deep learning system (DLS) compared to radiologists on (A) a combined dataset comprising 4 countries, (B) each dataset individually, (C) subgroups based on human immunodeficiency virus (HIV) status in the Zambia dataset, and (D) an additional test dataset from the mining population in South Africa.

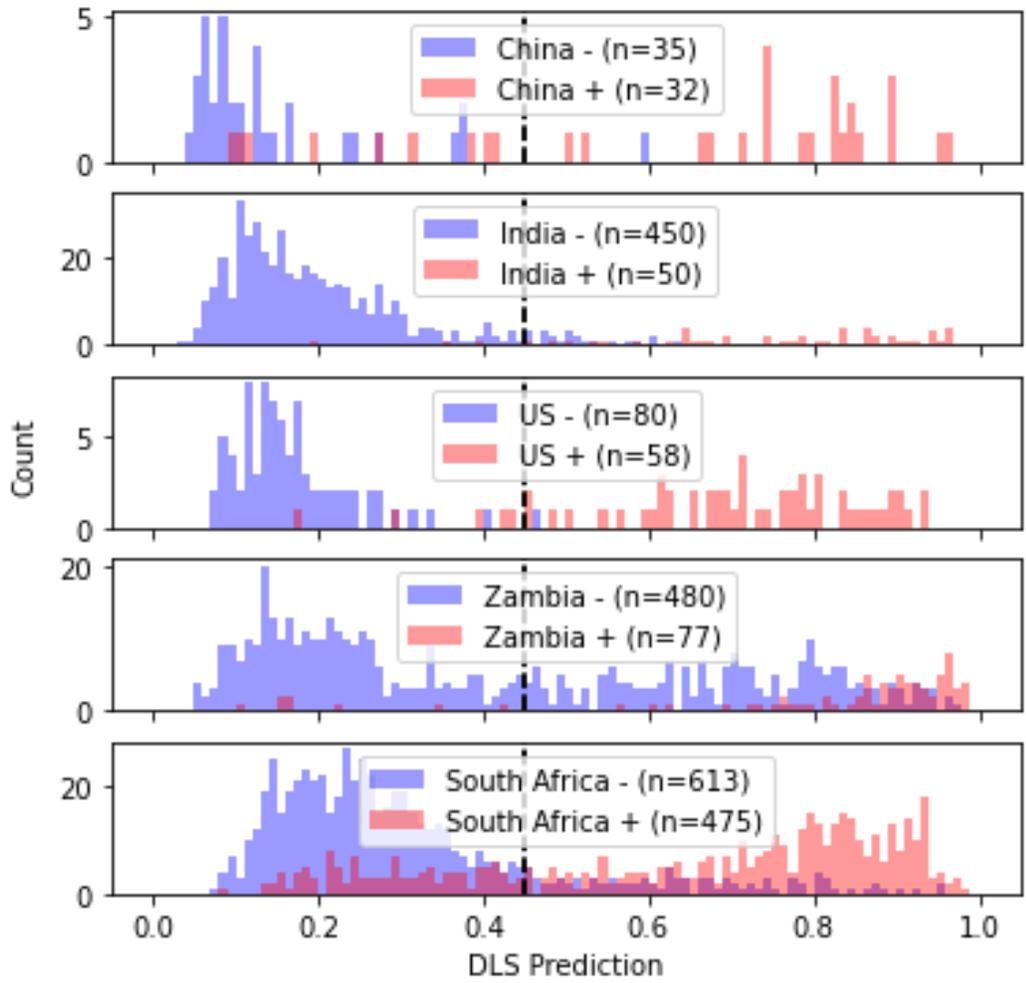

**Figure 3. Histograms representing the distribution of deep learning system (DLS) predictions stratified by positive (red) vs negative (blue) examples to illustrate shifts across datasets.**

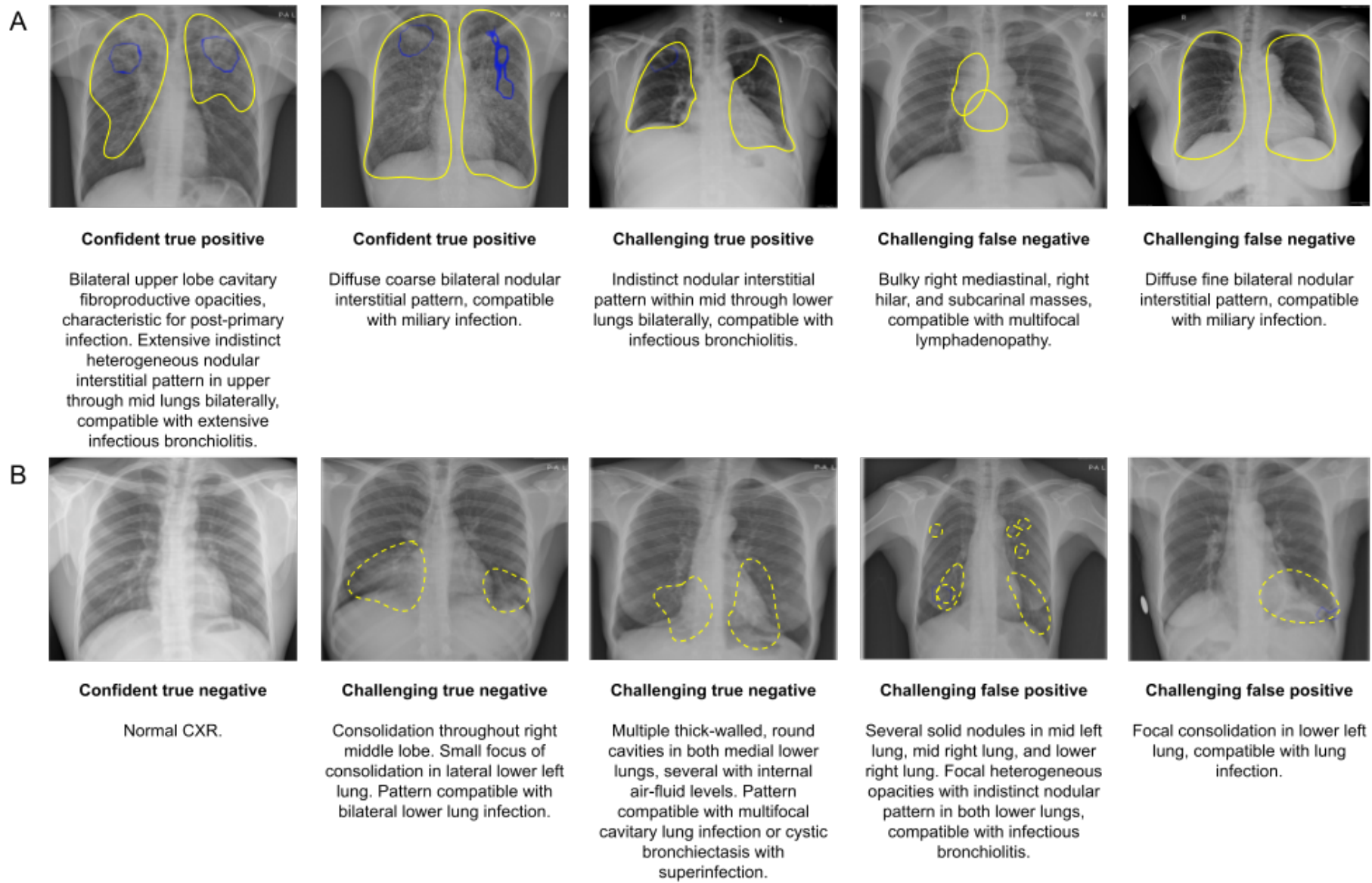

**Figure 4. Examples of chest radiographs (CXRs) corresponding to (A) TB-positive cases and (B) TB-negative cases.** Blue outlines encircle salient regions via Grad-CAM[47] that most influence the deep learning system (DLS) predictions, whereas yellow outlines were annotated by a radiologist to indicate regions of interest.

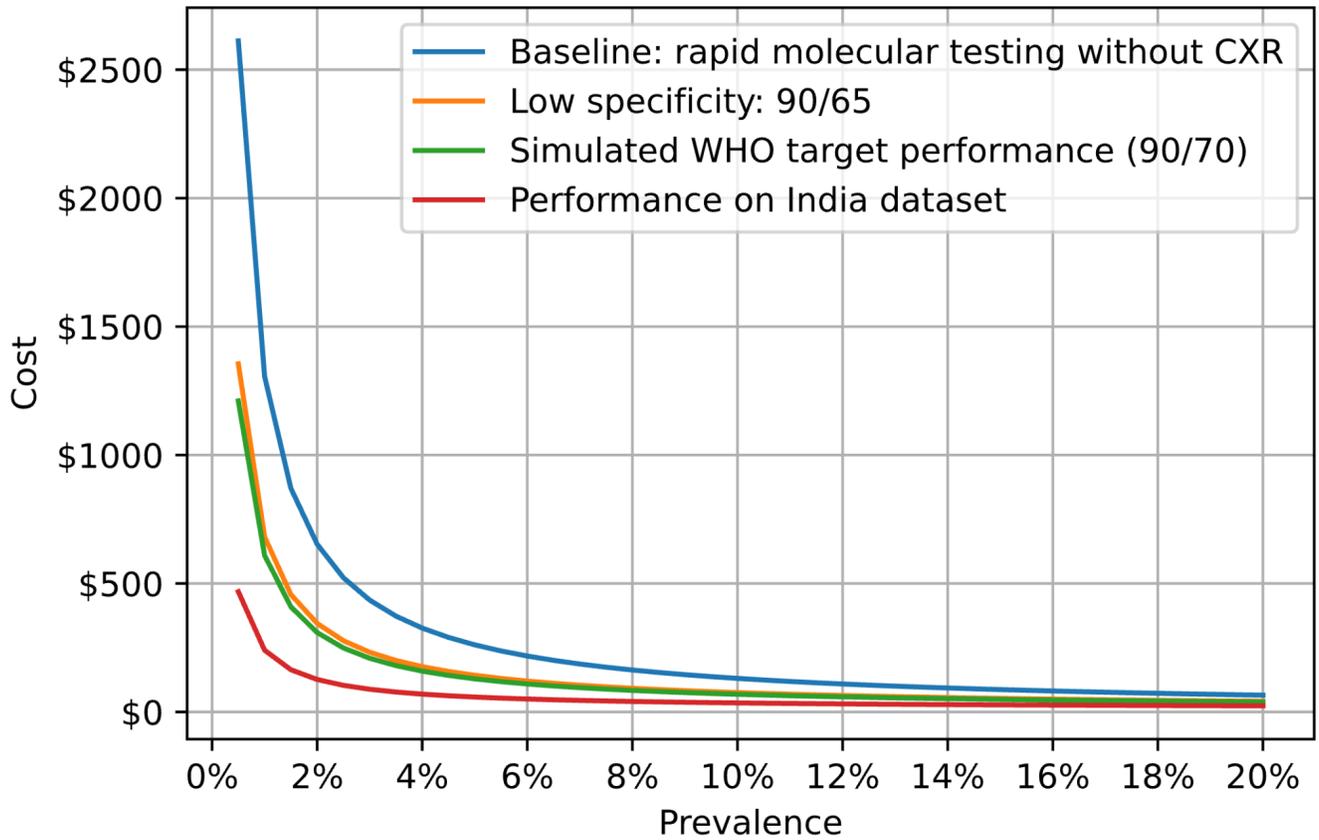

**Figure 5. The estimated cost per positive tuberculosis (TB) case caught using the deep learning system (DLS).** Absolute cost on the y-axis represents the expected cost per positive TB case detected.

# Tables

## Table 1. Baseline characteristics of datasets.

| | Train | | | | Tune | | | | Test | | | | Additional |
|---|---|---|---|---|---|---|---|---|---|---|---|---|---|
| | | | | | | | | | Combined | | | | |
| Geographical region | Europe (Azerbaijan, Belarus, Georgia, Moldova, Romania) and India | India | India | South Africa (Site A) | China (Shenzhen) | India | South Africa (Sites A,B,C) | Zambia (Lusaka) | China (Shenzhen) | India (Chennai) | US* (Montgomery, MD) | Zambia (Lusaka) | South Africa (Site D) |
| No. Patients | 822 | >268** | 2753 | 13270 | 595 | 466 | 496 | 1305 | 67 | 500 | 138 | 557 | 1088 |
| No. Female Patients (%) | 569 (0.69) | 134 (0.50) | 1793 (0.65) | 154 (0.01) | 405 (0.68) | 252 (0.54) | 6 (0.01) | 821 (0.63) | 44 (0.66) | 307 (0.61) | 19 (0.40) | 346 (0.62) | 21 (0.02) |
| No. Patients with Unknown Sex (%) | N.A. | N.A. | N.A. | 8035 (0.61) | N.A. | N.A. | 159 (0.32) | N.A. | N.A. | N.A. | N.A. | N.A. | 272 (0.25) |
| Race/Ethnicity | Predominantly Caucasian | Predominantly Indian | Predominantly Indian | Predominantly African | Predominantly East Asian | Predominantly Indian | Predominantly African | Predominantly African | Predominantly East Asian | Predominantly Indian | Predominantly Caucasian | Predominantly African | Predominantly African |
| Median Age | 40 | 49 | 53 | 42 | 32 | 47.5 | 41 | 34 | 34 | 47 | 40 | 32.5 | 42.5 |
| No. Images | 1096 | 5315 | 3266 | 150510 | 595 | 466 | 873 | 1324 | 67 | 500 | 138 | 557 | 1088 |
| No. Female Images | 749 (0.68) | 3040 (0.57) | 2143 (0.66) | 1702 (0.01) | 405 (0.68) | 252 (0.54) | 7 (0.01) | 834 (0.63) | 44 (0.66) | 307 (0.61) | 74 (0.54) | 347 (0.62) | 21 (0.02) |
| No. Images with Unknown Sex (%) | N.A. | N.A. | N.A. | 92094 (0.61) | N.A. | N.A. | 331 (0.38) | N.A. | N.A. | N.A. | N.A. | N.A. | 272 (0.25) |
| No. TB positive images | 1096 (1.00) | 1491 (0.28) | 1090 (0.33) | Unknown*** | 304 (0.51) | 134 (0.29) | 461 (0.53) | 189 (0.14) | 32 (0.48) | 50 (0.10) | 58 (0.42) | 77 (0.13) | 475 (0.44) |
| Reference Standard | Microbiology | Sputum culture | Radiologist | DLS Inferred | Mixed Clinical, Microbiological (Unconfirmed) | Sputum culture | Sputum culture, GeneXpert, Smear positive | GeneXpert | Mixed Clinical, Microbiological (Unconfirmed) | GeneXpert | Unconfirmed | GeneXpert | Sputum culture, GeneXpert, Smear positive |
| HIV status available | N.A. | N.A. | N.A. | 11233 | N.A. | N.A. | 217 | 1228 | N.A. | N.A. | N.A. | 525 | 232 |
| HIV positive (%) | N.A. | N.A. | N.A. | 8712 (0.78) | N.A. | N.A. | 182 | 474 (0.39) | N.A. | N.A. | N.A. | 217 (0.41) | 195 (0.84) |
| CXR View Available | 844 (77%) | N.A. | 2372 (73%) | 150354 (100%) | 595 (100%) | N.A. | 869 (99%) | 1324 (100%) | 67 (100%) | 413 (83%) | 138 (100%) | 562 (100%) | 1773 (99%) |
| CXR VIew PA | 835 (76%) | N.A. | 1892 (58%) | 137791 (92%) | 595 (100%) | N.A. | 789 (90%) | 1324 (100%) | 67 (100%) | 403 (81%) | 138 (100%) | 562 (100%) | 1762 (98%) |
| CXR VIew AP | 9 (1%) | N.A. | 480 (15%) | 12563 (8%) | 0 (0%) | N.A. | 80 (9%) | 0 (0%) | 0 (0%) | 10 (2%) | 0 (0%) | 0 (0%) | 11 (1%) |
| Image bit depth | 16 | 16 | 16 | 16 | 8 | 16 | 16 | 16 | 8 | 16 | 8 | 16 | 16 |
| Radiologist review for TB/no-TB | | | | | Yes 10 rads | Yes 4-5 | | Yes 20 rads | 10 India-based, 5 US-based | | | | |

*: The US dataset was not used for the TB classification model, but was used for developing the segmentation model (Methods). **: No. of unique patients not available; this is a lower bound based on the number of unique combinations of age, sex, and TB status. *** No TB test results available; the noisy student approach was used to infer labels for training purposes (Methods).

**Table 2. Comparing the DLS's sensitivity and specificity to radiologists.** Performance of the DLS at a second, prespecified high-sensitivity operating part is presented in Supplementary Table S4. Bold indicates p<0.05. *Prespecified 10% margin.

| Dataset | Comparator radiologists | Sensitivity | | | | | Specificity | | | | |
|---|---|---|---|---|---|---|---|---|---|---|---|
| | | Mean radiologist | DLS | Delta | Non-inferiority p-value* | Superiority p-value | Mean radiologist | DLS | Delta | Non-inferiority p-value* | Superiority p-value |
| Combined | India-based | 75.12% | 88.48% | **13.36%** | **<0.0001** | **<0.0001** | 83.99% | 78.66% | **-5.33%** | **0.0036** | 1.0000 |
| China | | 52.78% | 71.88% | **19.10%** | **0.0001** | **0.0051** | 95.87% | 97.14% | **1.27%** | **0.0001** | 0.3288 |
| India | | 84.89% | 94.00% | **9.11%** | **<0.0001** | **0.0120** | 96.77% | 95.11% | **-1.65%** | **<0.0001** | 1.0000 |
| US | | 65.13% | 91.38% | **26.25%** | **<0.0001** | **<0.0001** | 99.44% | 98.75% | **-0.69%** | **<0.0001** | 1.0000 |
| Zambia | | 85.57% | 89.61% | **4.04%** | **<0.0001** | **0.0106** | 68.56% | 58.54% | -10.02% | 0.5042 | 1.0000 |
| Combined | US-based | 78.34% | 88.48% | **10.14%** | **0.0215** | 0.1123 | 74.22% | 78.66% | **4.44%** | **0.0183** | 0.2019 |
| China | | 54.38% | 71.88% | **17.50%** | **0.0383** | 0.1100 | 93.71% | 97.14% | **3.43%** | **<0.0001** | 0.1202 |
| India | | 88.40% | 94.00% | **5.60%** | **0.0015** | 0.1238 | 94.40% | 95.11% | **0.71%** | **0.0007** | 0.3682 |
| US | | 73.10% | 91.38% | **18.28%** | **0.0071** | **0.0348** | 97.50% | 98.75% | **1.25%** | **<0.0001** | 0.2564 |
| Zambia | | 85.71% | 89.61% | 3.90% | 0.0976 | 0.3449 | 50.00% | 58.54% | 8.54% | 0.0543 | 0.2002 |

**Table 3. Comparing performance of the DLS to radiologists after matching to (A) mean radiologist specificity per dataset and (B) mean radiologist sensitivity per dataset.** Bold indicates p<0.05. *Prespecified 10% margin.

**A**

| Dataset | Comparator radiologists | Mean radiologist specificity | Mean radiologist's sensitivity | DLS's sensitivity | Delta in sensitivity | Non-inferiority p-value* | Superiority p-value |
|---|---|---|---|---|---|---|---|
| Combined | India-based | 83.99% | 75.12% | 82.95% | **7.83%** | **< 0.0001** | **0.0016** |
| China | | 95.87% | 52.78% | 81.25% | **28.47%** | **< 0.0001** | **0.0001** |
| India | | 96.77% | 84.89% | 90.00% | **5.11%** | **< 0.0001** | 0.0913 |
| US | | 99.44% | 65.13% | 87.93% | **22.80%** | **< 0.0001** | **< 0.0001** |
| Zambia | | 68.56% | 85.57% | 88.31% | **2.74%** | **< 0.0001** | 0.0556 |
| Combined | US-based | 74.22% | 78.34% | 92.17% | **13.82%** | **0.0125** | 0.0597 |
| China | | 93.71% | 54.38% | 81.25% | **26.88%** | **0.0148** | **0.0410** |
| India | | 94.40% | 88.40% | 94.00% | **5.60%** | **0.0015** | 0.1238 |
| US | | 97.50% | 73.10% | 96.55% | **23.45%** | **0.0035** | **0.0150** |
| Zambia | | 50.00% | 85.71% | 92.21% | 6.49% | 0.0683 | 0.2540 |

**B**

| Dataset | Comparator radiologists | Mean radiologist sensitivity | Mean radiologist's specificity | DLS's specificity | Delta in specificity | Non-inferiority p-value* | Superiority p-value |
|---|---|---|---|---|---|---|---|
| Combined | India-based | 75.12% | 83.99% | 87.75% | **3.76%** | **< 0.0001** | **0.0108** |
| China | | 52.78% | 95.87% | 100.00% | **4.13%** | **< 0.0001** | **0.0180** |
| India | | 84.89% | 96.77% | 98.67% | **1.90%** | **< 0.0001** | 0.0977 |
| US | | 65.13% | 99.44% | 100.00% | **0.56%** | **< 0.0001** | 0.1040 |
| Zambia | | 85.57% | 68.56% | 76.46% | **7.89%** | **< 0.0001** | **0.0008** |
| Combined | US-based | 78.34% | 74.22% | 86.70% | **12.48%** | **0.0041** | **0.0282** |
| China | | 54.38% | 93.71% | 100.00% | **6.29%** | **< 0.0001** | **0.0209** |

| | | | | | | |
|---|---|---|---|---|---|---|
| India | | 88.40% | 94.40% | 97.11% | **2.71%** | **0.0004** | 0.1076 |
| US | | 73.10% | 97.50% | 100.00% | **2.50%** | **0.0004** | 0.0710 |
| Zambia | | 85.71% | 50.00% | 76.46% | **26.46%** | **0.0073** | **0.0207** |

**Table 4. Model performance at the World Health Organization's (WHO) target sensitivity and specificity thresholds.**

| Dataset | DLS Specificity @ 90% Sensitivity | DLS Sensitivity @ 70% Specificity |
|---|---|---|
| Combined | 76.65% | 93.55% |
| China | 80.00% | 90.63% |
| India | 97.11% | 98.00% |
| US | 98.75% | 98.28% |
| Zambia | 56.25% | 87.01% |

# Supplementary Figures

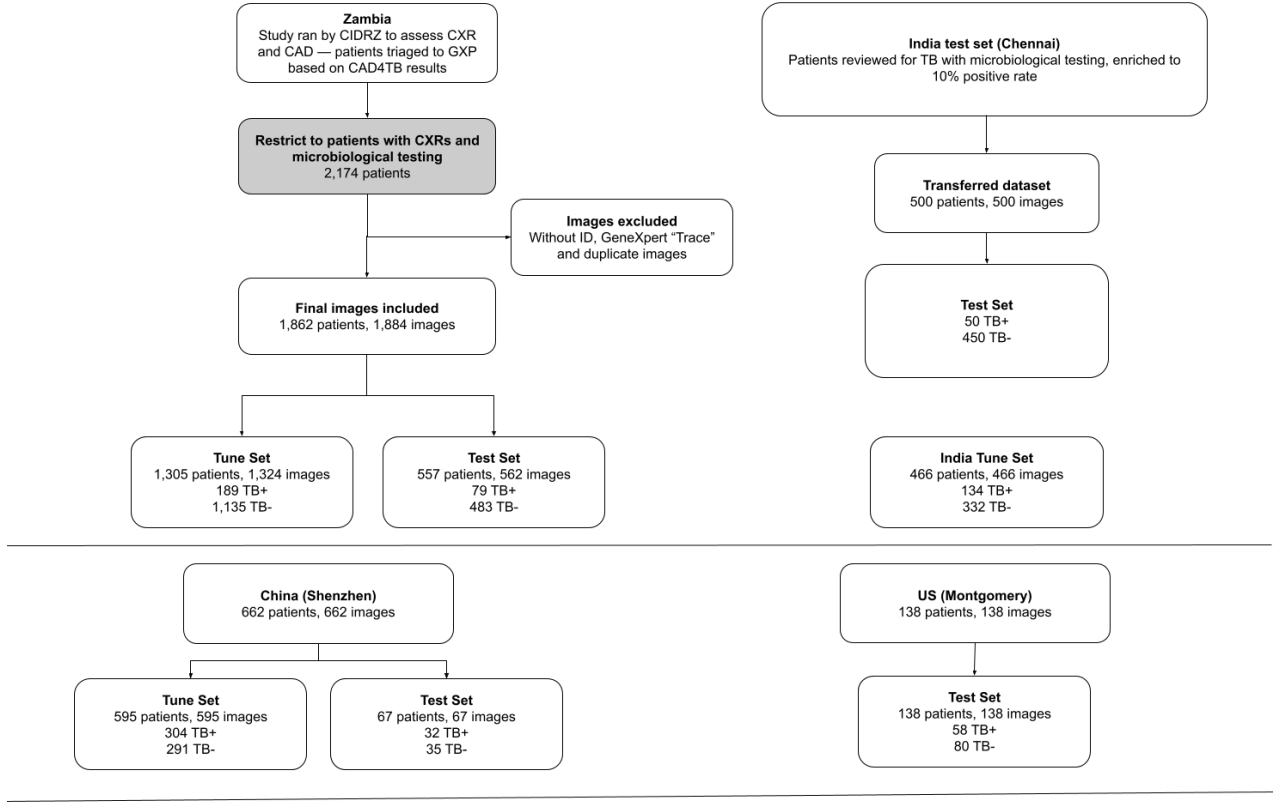

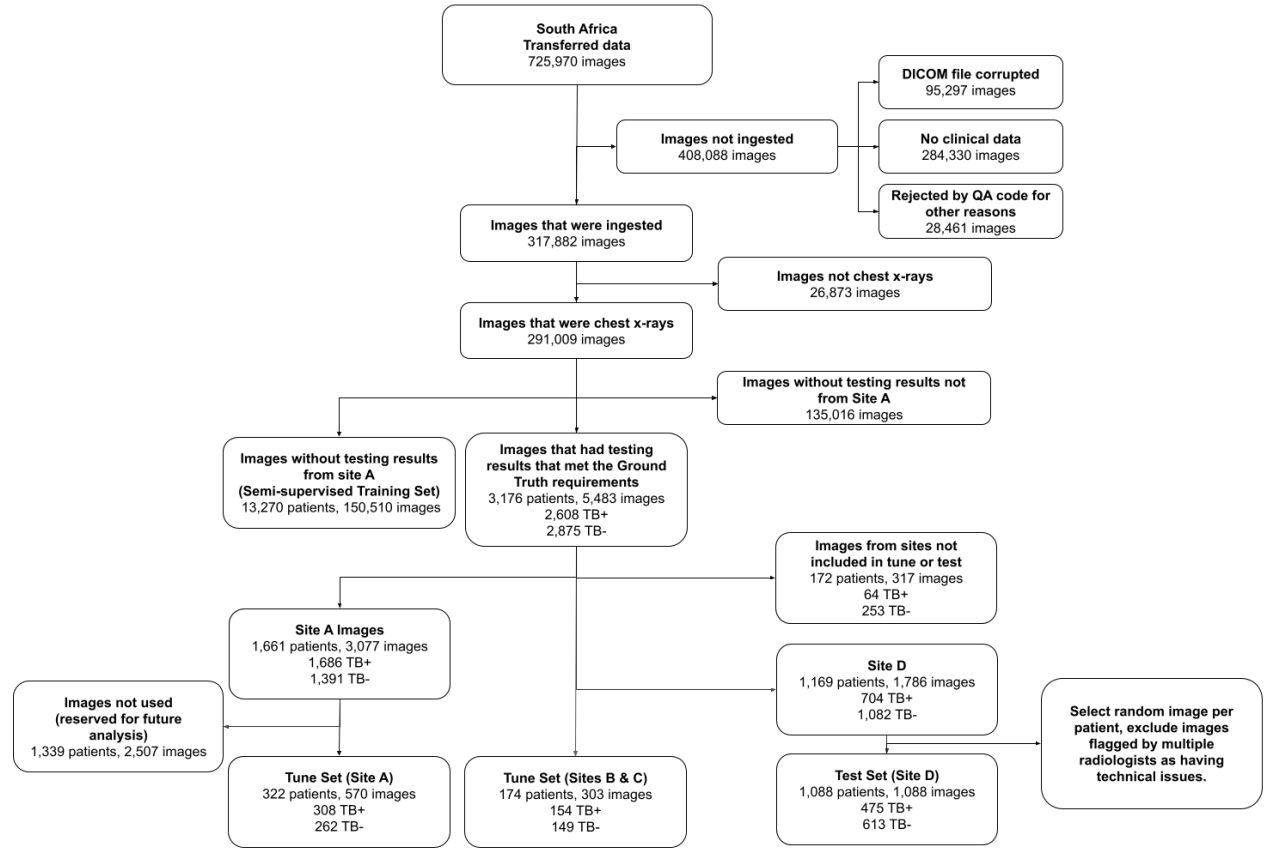

**Supplementary Figure S1. STARD diagrams.**

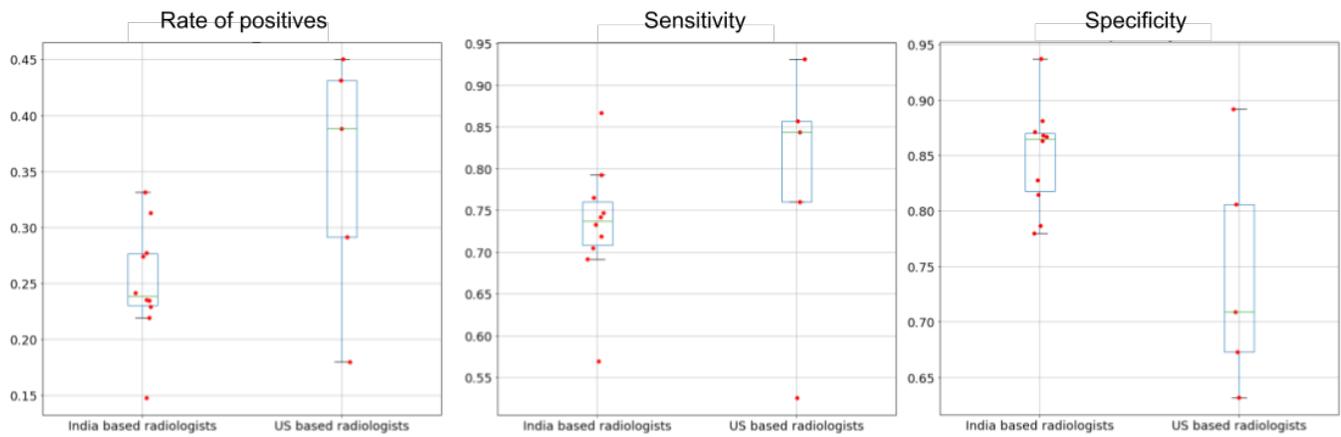

**Supplementary Figure S2. Boxplots showing the rates of flagging images as positive per grader.** The outlier India-based radiologist was excluded owing to a substantially lower rate of flagging positives (15% vs 22-33%) and accordingly an anomalously-low sensitivity (57% vs 69-86%). Outliers were defined using matplotlib default settings, as point beyond the whiskers (1.5 * interquartile range away from the first or third quartiles).

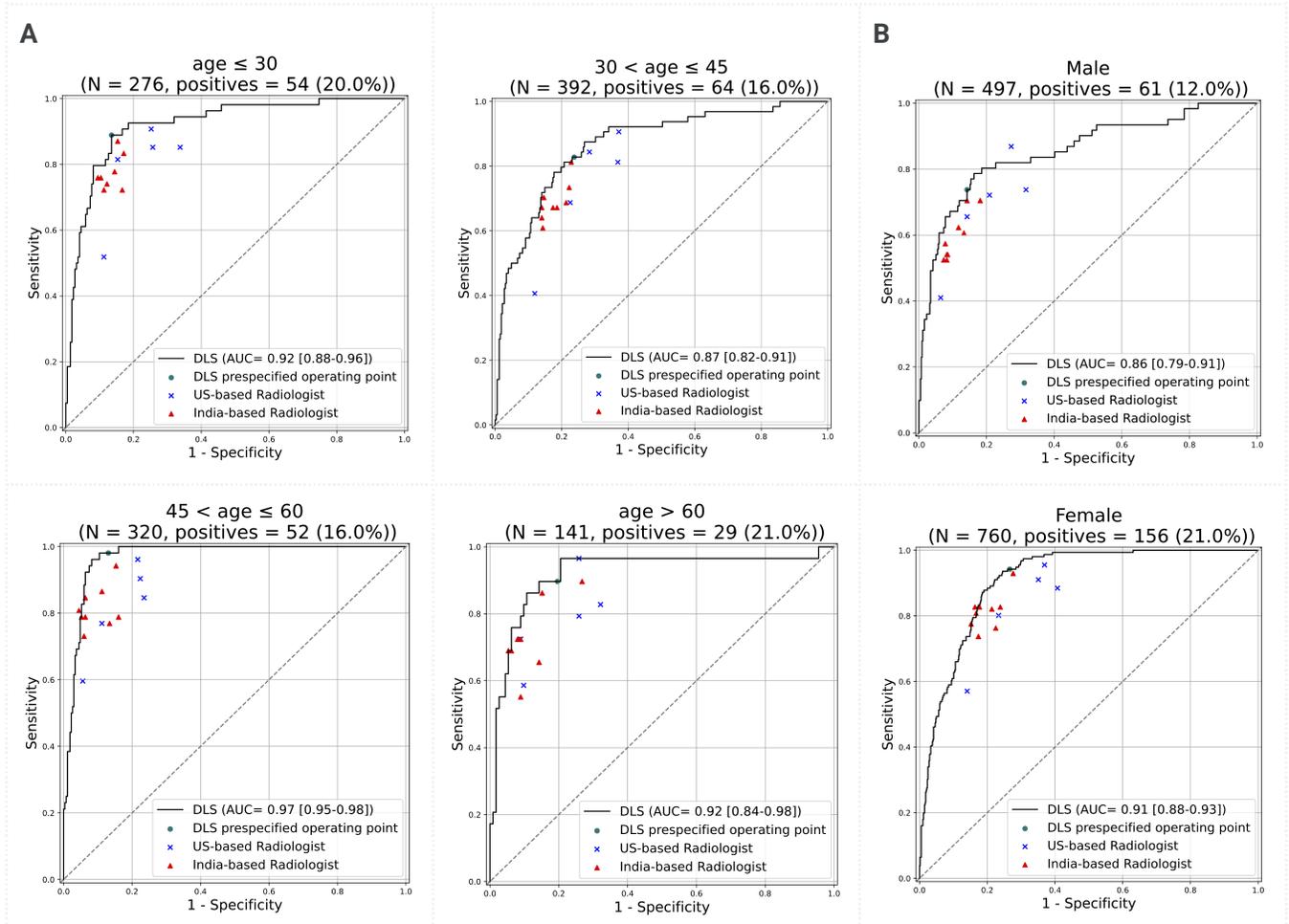

**Supplementary Figure S3.** Subgroups based on (A) age and (B) sex for the combined test set across China, India, US, and Zambia.

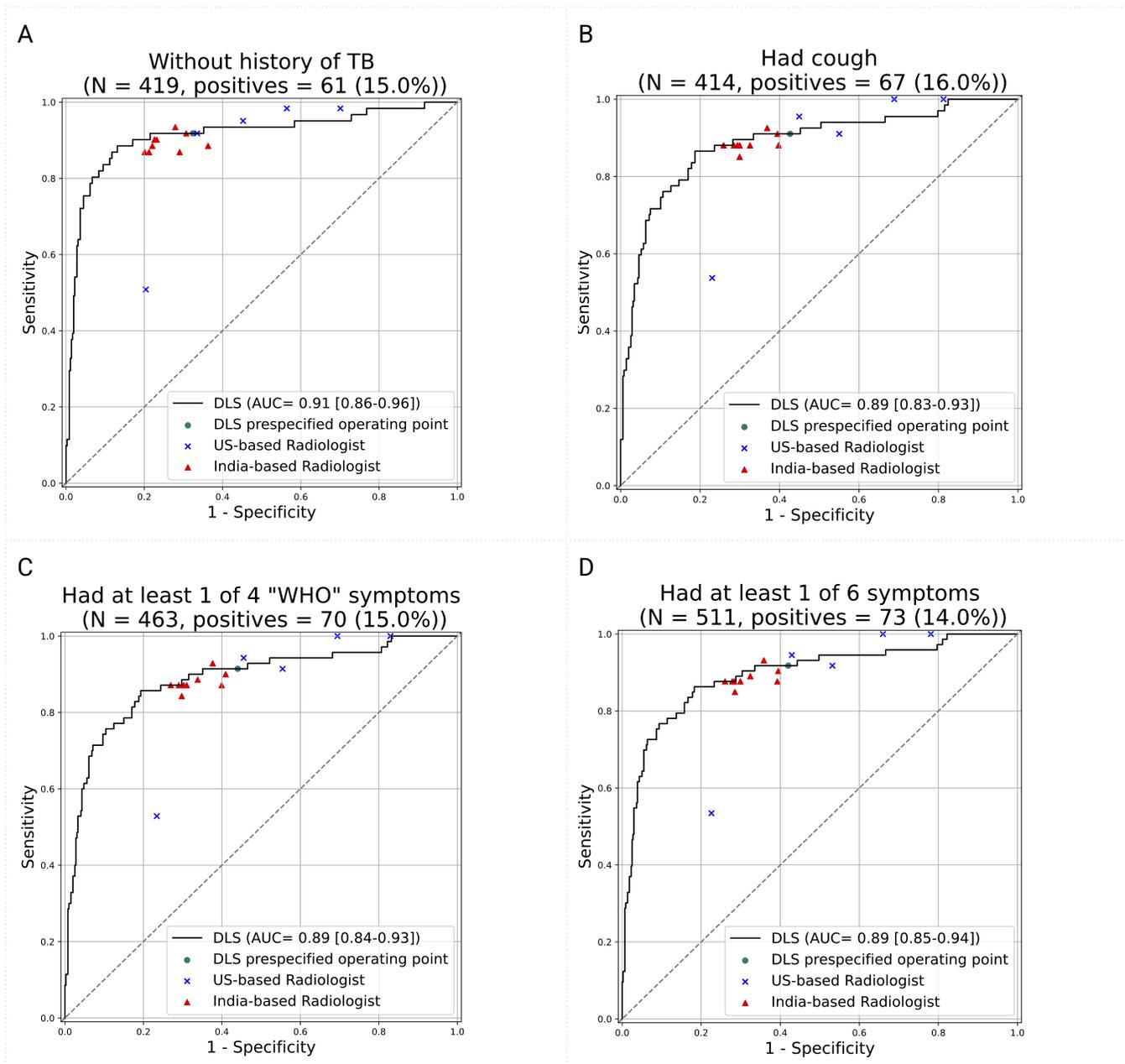

**Supplementary Figure S4.** Subgroups based on (A) having no history of tuberculosis, (B) having a cough, (C) having any one of WHO's four symptom symptom screen (cough, weight loss, fever, or night sweats), and (D) having any of 6 symptoms: cough, weight loss, fever, night sweats, shortness of breath, or chest pain.

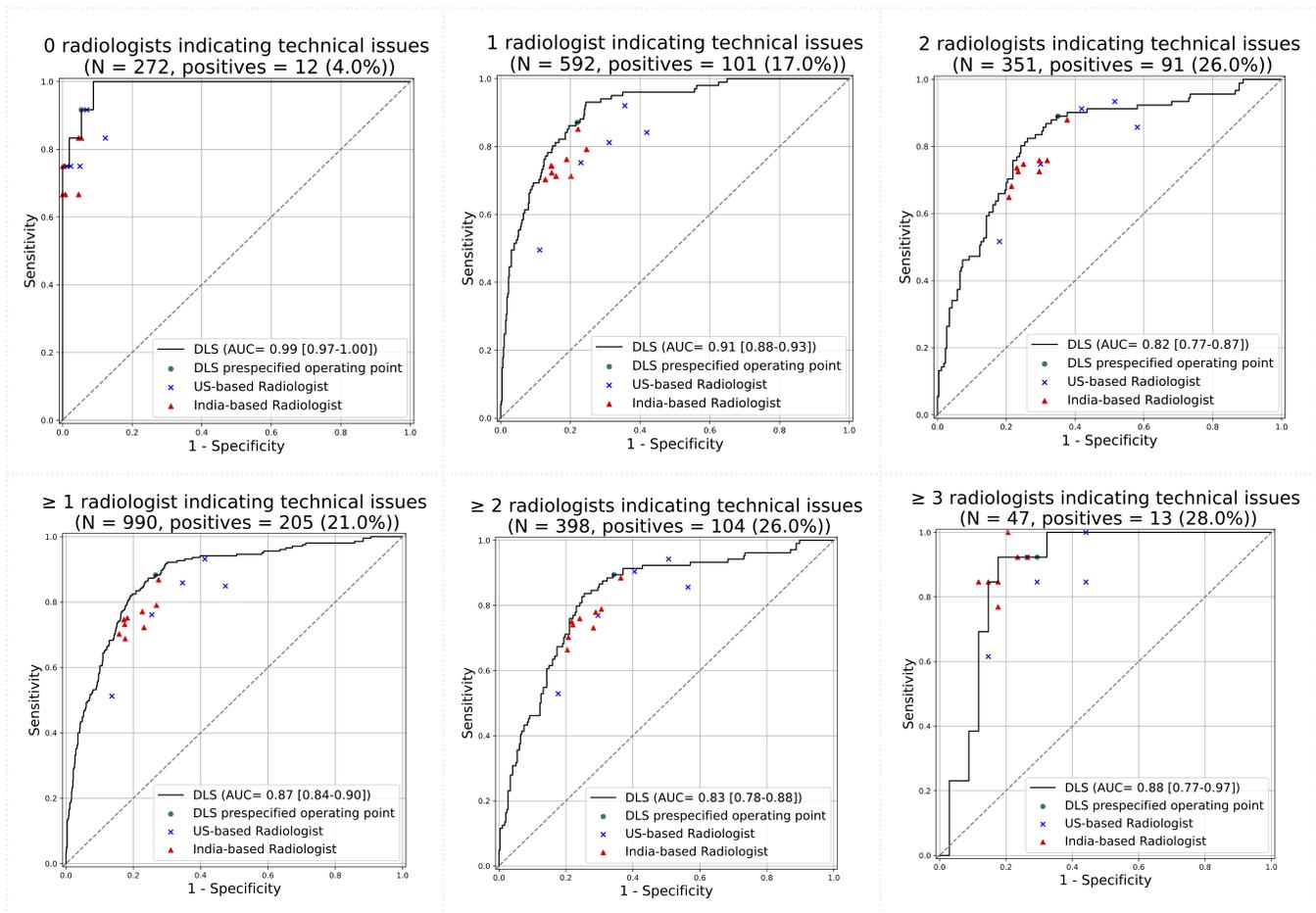

**Supplementary Figure S5.** Subgroups based on the number of radiologists who indicated the presence of a minor technical issue: 0, 1, 2, and ≥3, as well as a cumulative range: ≥1 and ≥2.

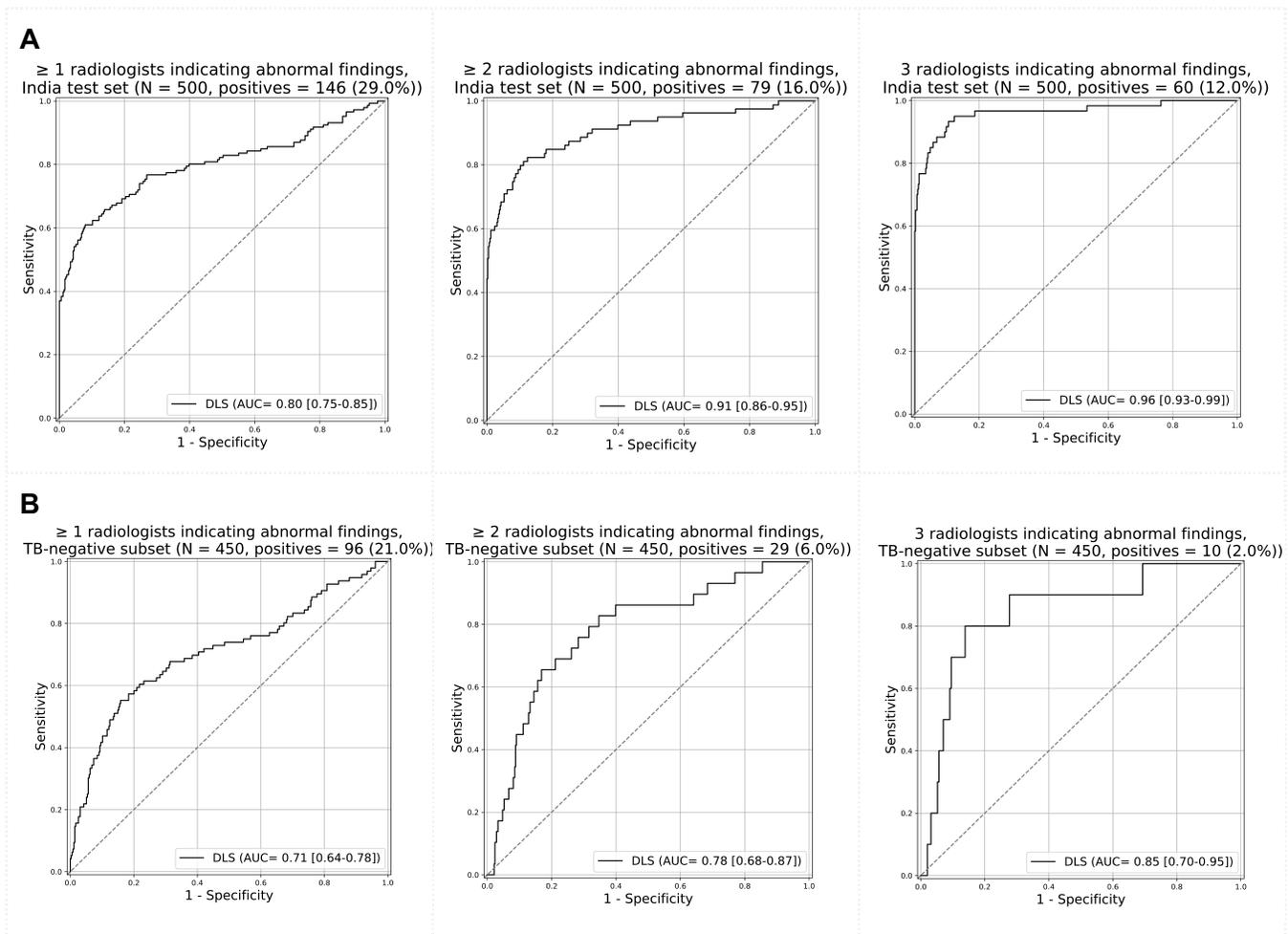

**Supplementary Figure S6. Performance of the DLS for detecting abnormalities in (A) the entire India test set, and (B) the TB-negative examples in the India test set.** The reference standard for whether abnormalities are present are based on US-based radiologists' response to whether there were any abnormal findings present in the image. From left to right, at least 1, 2, or 3 (out of 3) radiologists indicated the image was positive for such findings, indicating additional certainty in the positive label. In (A), a positive case is defined as either being TB positive or having radiologists indicating presence of other findings; the DLS's positive prediction is the sum of its TB prediction and abnormality prediction. In (B), all TB-positive cases have been removed; a positive case is defined via radiologists indicating presence of other findings; the DLS's positive prediction is its abnormality prediction alone.

# Supplementary Tables

## Supplementary Table S1. Detailed dataset information.

| Dataset / Location | Description | Tests | Label Criteria |
|---|---|---|---|
| South Africa (various locations) | Collected from periodic employee screening provided by a Gold Mining Company. Patients suspected of having TB based on clinical assessment and radiographic evidence receive GeneXpert and/or culture testing. | GeneXpert Culture Smear | Positive: <br> - Positive TB test (GeneXpert, Culture, Smear) within +/- 45 days of CXR. <br> - No negative GeneXpert or sputum culture within +/- 150 days of CXR. <br> Negative: <br> - Negative GeneXpert or sputum culture test within +/- 150 days of the CXR. <br> - Never registered in the company TB register. |
| Lusaka, Zambia | This dataset includes patients who were enrolled in a TB outreach study involving case finding at healthcare facilities and in the community. When CXR was available, all patients were screened with CAD4TBv5 followed by GeneXpert for patients with abnormal CXRs and at the discretion of the clinician. When a CXR was not available, symptomatic patients were asked to provide sputum for GeneXpert. Source | GeneXpert Culture Smear | Positive: <br> - Positive TB test (GeneXpert, Culture, Smear) <br> Negative: <br> - Negative TB test (GeneXpert, Culture) <br><br> Excluded: <br> - People screened by CAD4TBv5 who did not get a GeneXpert or Culture. <br> - Patient with TB trace results from GeneXpert test |
| Shenzhen, China | This public dataset includes a curated subset TB case-control patients who presented to a Shenzhen Hospital for routine care. Source | Not described | Positive: <br> - Positive TB test (eg, molecular or culture) <br> Negative: <br> - Radiologically clear or negative culture |
| India | **Private tertiary Indian hospital:** <br> This dataset includes patients who presented to the tertiary hospital with symptoms suggestive of TB. <br><br> **Primary health clinics:** <br> This dataset includes patients who presented to community clinics with symptoms suggestive of TB. | NAAT | Positive: <br> - Positive rapid molecular testing <br> Negative: <br> - Patients without clinical suspicion of TB or having negative rapid molecular testing |
| Montgomery, Maryland, US | This public dataset includes a curated subset TB case-control patients who were part of Montgomery County's Tuberculosis screening program. | Culture | Positive: <br> - Positive TB test (eg molecular or culture) <br> Negative: <br> - Confirmation not documented |
| Chennai, India | This dataset includes patients referred to a private Indian hospital for TB testing based on symptoms or exposure. | NAAT | Positive: <br> - Positive rapid molecular testing <br> Negative: <br> - Negative rapid molecular testing |

**Supplementary Table S2. The DLS's sensitivity and specificity compared to radiologists in the South Africa dataset.** Two operating points are shown here: the same prespecified operating point as the other datasets (A: 0.45), and a second operating point (B: 0.685). Note that 1 India-based radiologist was excluded in the presented analysis as they did not complete all of the cases due to a technical issue. Bold indicates p<0.05. *Prespecified 10% margin.

| Operating point | Comparator radiologists | Sensitivity | | | | | Specificity | | | | |
| --- | --- | --- | --- | --- | --- | --- | --- | --- | --- | --- | --- |
| | | Mean radiologist | DLS | Delta | Non-inferiority p-value* | Superiority p-value | Mean radiologist | DLS | Delta | Non-inferiority p-value* | Superiority p-value |
| A | India-based | 55.68% | 75.16% | **19.47%** | **<0.0001** | **0.0005** | 85.48% | 84.67% | **-0.82%** | **0.0061** | 1.0000 |
| B | | 55.68% | 56.21% | **0.53%** | **0.0097** | 0.8908 | 85.48% | 94.29% | **8.81%** | **0.0001** | **0.0148** |
| A | US-based | 67.66% | 75.16% | **7.49%** | **0.0404** | 0.3800 | 79.87% | 84.67% | **4.80%** | **0.0050** | 0.2463 |
| B | | 67.66% | 56.21% | -11.45% | 0.5705 | 1.0000 | 79.87% | 94.29% | **14.42%** | **0.0006** | **0.0115** |

**Supplementary Table S3. Comparing sensitivity of AI and radiologists at (A) matching specificity and (B) matching sensitivity to individual radiologists in the external test set (Chennai, India) and a split-sample test set (Zambia).** This analysis compared the AI with each individual radiologist. P-values are based on the Wald test for non-inferiority and McNemar test for superiority. Bold indicates p<0.05. *Prespecified 10% margin.

**A**

| Dataset | Radiologist # | Radiologist specificity (%) | Radiologist sensitivity (%) | AI sensitivity | Delta in sensitivity | Non-inferiority p-value* | Superiority p-value |
|---|---|---|---|---|---|---|---|
| India: N=500 cases; n=50 TB-positive (10%) | India 1 | 91.11% | 86.00% | 96.00% | **10.00%** | **< 0.001** | 0.059 |
| | India 2 | 99.56% | 84.00% | 82.00% | **-2.00%** | **0.037** | 1.000 |
| | India 3 | 98.89% | 84.00% | 84.00% | **0.00%** | **0.021** | 1.000 |
| | India 4 | 99.56% | 84.00% | 82.00% | -2.00% | 0.065 | 1.000 |
| | India 5 | 98.22% | 84.00% | 88.00% | **4.00%** | **< 0.001** | 0.317 |
| | India 6 | 91.33% | 94.00% | 96.00% | **2.00%** | **0.004** | 0.655 |
| | India 7 | 92.89% | 86.00% | 94.00% | **8.00%** | **0.002** | 0.206 |
| | India 8 | 100.00% | 80.00% | 80.00% | **0.00%** | **0.021** | 1.000 |
| | India 9 | 99.33% | 82.00% | 82.00% | **0.00%** | **0.021** | 1.000 |
| | ~~India 10~~* | 99.56% | 58.00% | 82.00% | **24.00%** | **< 0.001** | **< 0.001** |
| | US 1 | 92.44% | 94.00% | 94.00% | **0.00%** | **0.006** | 1.000 |
| | US 2 | 93.78% | 90.00% | 94.00% | **4.00%** | **0.006** | 0.480 |
| | US 3 | 97.78% | 86.00% | 88.00% | **2.00%** | **0.012** | 0.705 |
| | US 4 | 98.89% | 78.00% | 84.00% | **6.00%** | **0.004** | 0.317 |
| | US 5 | 89.11% | 94.00% | 98.00% | **4.00%** | **< 0.001** | 0.157 |
| Zambia: N=557 cases; n=77 TB-positive (14%) | India 1 | 68.75% | 84.42% | 88.31% | **3.90%** | **< 0.001** | 0.257 |
| | India 2 | 72.50% | 85.71% | 85.71% | **0.00%** | **< 0.001** | 1.000 |
| | India 3 | 72.50% | 84.42% | 85.71% | **1.30%** | **0.002** | 0.739 |
| | India 4 | 71.67% | 81.82% | 85.71% | **3.90%** | **< 0.001** | 0.180 |
| | India 5 | 64.58% | 90.91% | 89.61% | **-1.30%** | **< 0.001** | 1.000 |
| | India 6 | 60.83% | 88.31% | 89.61% | **1.30%** | **< 0.001** | 0.564 |
| | India 7 | 61.25% | 85.71% | 89.61% | **3.90%** | **< 0.001** | 0.083 |
| | India 8 | 70.21% | 85.71% | 87.01% | **1.30%** | **< 0.001** | 0.317 |
| | India 9 | 74.79% | 83.12% | 85.71% | **2.60%** | **< 0.001** | 0.414 |
| | ~~India 10~~* | 86.67% | 64.94% | 75.32% | **10.39%** | **< 0.001** | 0.059 |
| | US 1 | 36.88% | 97.40% | 92.21% | **-5.19%** | **0.029** | 1.000 |
| | US 2 | 26.67% | 97.40% | 93.51% | **-3.90%** | **0.003** | 1.000 |
| | US 3 | 78.75% | 50.65% | 84.42% | **33.77%** | **< 0.001** | **< 0.001** |
| | US 4 | 58.96% | 92.21% | 89.61% | **-2.60%** | **0.010** | 1.000 |
| | US 5 | 48.75% | 90.91% | 92.21% | **1.30%** | **< 0.001** | 0.564 |

*This radiologist was an outlier and excluded in the other analyses (Supplementary Figure S2); presented here for completeness.

**B**

| Dataset | Radiologist # | Radiologist specificity (%) | Radiologist sensitivity (%) | AI sensitivity | Delta in sensitivity | Non-inferiority p-value* | Superiority p-value |
|---|---|---|---|---|---|---|---|
| India: N=500 cases; n=50 TB-positive (10%) | India 1 | 86.00% | 91.11% | 98.67% | **7.56%** | **< 0.001** | **< 0.001** |
| | India 2 | 84.00% | 99.56% | 99.11% | **-0.44%** | **< 0.001** | 1.000 |
| | India 3 | 84.00% | 98.89% | 99.11% | **0.22%** | **< 0.001** | 0.655 |
| | India 4 | 84.00% | 99.56% | 99.11% | **-0.44%** | **< 0.001** | 1.000 |
| | India 5 | 84.00% | 98.22% | 99.11% | **0.89%** | **< 0.001** | 0.248 |
| | India 6 | 94.00% | 91.33% | 95.56% | **4.22%** | **< 0.001** | **0.005** |
| | India 7 | 86.00% | 92.89% | 98.67% | **5.78%** | **< 0.001** | **< 0.001** |
| | India 8 | 80.00% | 100.00% | 100.00% | **0.00%** | **< 0.001** | 1.000 |
| | India 9 | 82.00% | 99.33% | 99.78% | **0.44%** | **< 0.001** | 0.317 |
| | ~~India 10*~~ | 58.00% | 99.56% | 100.00% | **0.44%** | **< 0.001** | 0.157 |
| | US 1 | 94.00% | 92.44% | 95.56% | **3.11%** | **< 0.001** | **0.035** |
| | US 2 | 90.00% | 93.78% | 97.11% | **3.33%** | **< 0.001** | **0.009** |
| | US 3 | 86.00% | 97.78% | 98.67% | **0.89%** | **< 0.001** | 0.317 |
| | US 4 | 78.00% | 98.89% | 100.00% | **1.11%** | **< 0.001** | **0.025** |
| | US 5 | 94.00% | 89.11% | 95.56% | **6.44%** | **< 0.001** | **< 0.001** |
| Zambia: N=557 cases; n=77 TB-positive (14%) | India 1 | 84.42% | 68.75% | 81.04% | **12.29%** | **< 0.001** | **< 0.001** |
| | India 2 | 85.71% | 72.50% | 76.46% | **3.96%** | **< 0.001** | **0.017** |
| | India 3 | 84.42% | 72.50% | 81.04% | **8.54%** | **< 0.001** | **< 0.001** |
| | India 4 | 81.82% | 71.67% | 82.50% | **10.83%** | **< 0.001** | **< 0.001** |
| | India 5 | 90.91% | 64.58% | 56.25% | **-8.33%** | 0.177 | 1.000 |
| | India 6 | 88.31% | 60.83% | 69.58% | **8.75%** | **< 0.001** | **< 0.001** |
| | India 7 | 85.71% | 61.25% | 76.46% | **15.21%** | **< 0.001** | **< 0.001** |
| | India 8 | 85.71% | 70.21% | 76.46% | **6.25%** | **< 0.001** | **< 0.001** |
| | India 9 | 83.12% | 74.79% | 81.46% | **6.67%** | **< 0.001** | **< 0.001** |
| | ~~India 10*~~ | 64.94% | 86.67% | 94.17% | **7.50%** | **< 0.001** | **< 0.001** |
| | US 1 | 97.40% | 36.88% | 18.96% | -17.92% | 1.000 | 1.000 |
| | US 2 | 97.40% | 26.67% | 18.96% | -7.71% | 0.166 | 1.000 |
| | US 3 | 50.65% | 78.75% | 96.46% | **17.71%** | **< 0.001** | **< 0.001** |
| | US 4 | 92.21% | 58.96% | 51.25% | -7.71% | 0.136 | 1.000 |
| | US 5 | 90.91% | 48.75% | 56.25% | **7.50%** | **< 0.001** | **< 0.001** |

*This radiologist was an outlier and excluded in the other analyses (Supplementary Figure S2); presented here for completeness.

**Supplementary Table S4. Performance of the DLS at a prespecified high sensitivity operating point.**

| Dataset | Sensitivity | Specificity |
|---|---|---|
| Combined | 94.5% | 61.8% |
| Shenzhen | 87.5% | 85.7% |
| India | 98.0% | 77.3% |
| Montgomery | 98.3% | 88.7% |
| Zambia | 92.2% | 41.0% |

**Supplementary Table S5. Numerical comparisons across the 4 test datasets in terms of DLS predictions for (A) all cases, (B) TB-positive cases, and (C) TB-negatives.** In the header row, numbers represent mean DLS prediction ± standard deviation. Each cell indicates the p-value for the difference in the distributions via the Kolmogorov-Smirnov test.

**A**. All cases

|  | China (n=67) | India (n=500) | US (n=138) | Zambia (n=557) |
|---|---|---|---|---|
| China (n=67) |  |  |  |  |
| India (n=500) | 0.0002 |  |  |  |
| US (n=138) | 0.0082 | <0.0001 |  |  |
| Zambia (n=557) | <0.0001 | <0.0001 | 0.0002 |  |
| South Africa (n=1088) | <0.0001 | <0.0001 | <0.0001 | 0.0043 |

**B**. TB-positive cases

|  | China (n=32) | India (n=50) | US (n=58) | Zambia (n=77) |
|---|---|---|---|---|
| China (n=32) |  |  |  |  |
| India (n=50) | 0.2484 |  |  |  |
| US (n=58) | 0.1977 | 0.0736 |  |  |
| Zambia (n=77) | <0.0001 | 0.0014 | <0.0001 |  |
| South Africa (n=475) | 0.9538 | 0.0481 | 0.0622 | <0.0001 |

**C**. TB-negative cases

|  | China (n=35) | India (n=450) | US (n=80) | Zambia (n=480) |
|---|---|---|---|---|
| China (n=35) |  |  |  |  |
| India (n=450) | <0.0001 |  |  |  |
| US (n=80) | 0.0002 | 0.0288 |  |  |
| Zambia (n=480) | <0.0001 | <0.0001 | <0.0001 |  |
| South Africa (n=613) | <0.0001 | <0.0001 | <0.0001 | <0.0001 |